\documentstyle[pra,aps,amssymb,epsf]{revtex}

\newcommand{\bra}[1]{\mathop{\left\langle #1 \right|}\nolimits}
\newcommand{\ket}[1]{\mathop{\left| #1 \right\rangle}\nolimits}

\newcommand{\avr}[1]{\mathop{\left\langle #1\right\rangle}\nolimits}

\begin{document}
\draft

\title{Natural Capacity of a System of Two Two-Level Atoms\\
as a Quantum Information Channel}

\author{B.\ A.\ Grishanin and V.\ N.\ Zadkov}
\address{Physical Faculty and International Laser Center,
M.\ V.\ Lomonosov Moscow State University Moscow, 119899 Russia}
\maketitle

\begin{abstract}
A system of two closely spaced atoms interacting through a vacuum
electromagnetic field is considered. It is demonstrated that
radiative decay in such a system resulting from photon exchange
gives rise to a definite amount of information related to
interatomic communication. Joint distributions of detection
probabilities of atomic quanta and the corresponding amount of
communication information are calculated.
\end{abstract}

\section{Introduction}
\label{intro:}

Analysis of physical systems as potential sources of quantum
information is becoming an urgent issue in the context of extensive
research into the physical implementation of quantum computation
techniques [1-3]. Such an analysis is inevitably associated with
mathematical problems that differ from those encountered in the
analysis of such systems as objects of conventional methods of
physical experiments. One of the natural goals of the approach
specified above is to reveal the possibilities of using specific
physical mechanisms that would ensure information exchange between
microscopic quantum systems employed as components of quantum data
converters. In this respect, two-level atoms (TLAs) interacting
through a vacuum electro-magnetic field can be considered as one of
the fundamental systems of this type. Obviously, two two-level
atoms (TTLAs) form an elementary system. Certain efforts have been
already concentrated on the investigation of such a system [4-7].
The main problem encountered in the exact calculation of TTLA
dynamics stems from the necessity to rigorously take into account
relaxation processes simultaneously with reversible interactions.

In this paper, we investigate the process of radiative decay in a
TTLA system in its pure form in the absence of an external field.
We consider TTLA dynamics on a time scale $\tau\gg\tau_r$, i.e.,
for time intervals greater than the radiative decay time $\tau_r$
for a single atom. From the viewpoint of information transmission
in a TTLA system, it is of interest to understand how much
information is produced and stored in a system upon the completion
of radiative decay processes in TLAs constituting the system under
study and which type of information we deal with in this case.
Evidently, such a formulation of the problem has no meaning for
atoms separated by a distance on the order of or greater than the
wavelength, i.e., for atoms in traps, which can be considered as
one of the prototypes of a physical quantum processor [3]. In this
case, all the dynamic variables of a TTLA system decay on the same
time scale $\tau_r$. However, for atoms of the relevant variables
continue to relax within much greater time intervals
$\tau\gg\tau_r.$ Then, the information that relates the initial
state of a TTLA system to its final state is stored. Such a
geometry of a TTLA system may be implemented not only for atoms in
dense media, but is also typical of impurity atoms adsorbed on a
substrate. According to the experimental data of [8], impurity
atoms under these conditions may preserve the discrete structure of
atomic states and may be considered as microscopic quantum systems
for quantum data con-version in a quantum computer. The mechanism
behind the information exchange described above is associated with
a relaxation photon exchange between closely spaced TLAs initially
prepared in independent states. If the exchange rate $\gamma_{\rm
ex}$ is close to the radiative decay rate $\gamma$ of TLAs, then a
single-quantum diatomic state $\psi_{\rm a}=\ket{1}
\otimes\ket{0} -\ket{0} \otimes \ket{1}$, which decays with a rate
equal to $\gamma-\gamma_{\rm ex}$, exists within a time interval
$\tau\gg\tau_r=\gamma^{-1}$. It is this singlet (in terms of a
system of two spins) state that plays an important role in systems
implementing the methods of quantum cryptography [9, 10]. The
mechanism under consideration naturally governs quantum data
exchange between TLAs.

In this paper, we restrict our consideration to a system of two
closely spaced TLAs with a geometry of dipole moments shown in Fig.
1. Obviously, as long as the application of such a system is not
specified, its information efficiency remains uncertain.
Nevertheless, it seems appropriate to define the information
efficiency in terms of the natural definition of the amount of
data, which will be specified in this paper with the use of a given
procedure of quantum measurement, without discussing the
generalization of the considered approach. We assume that the
information criterion employed in our study is sufficient for
comparing different mechanisms of information exchange in terms of
the information efficiency of feasible schemes for quantum
computations using atomic transitions. In this paper, the
definition of the amount of information will employ the Shannon
definition of communication information corresponding to the
probability distribution of measured energy quanta for each atom,
which is described by a standard expression presented in [11].

\begin{figure}[bht]
\begin{center}
\epsfysize=4.cm\epsfclipon\leavevmode\epsffile{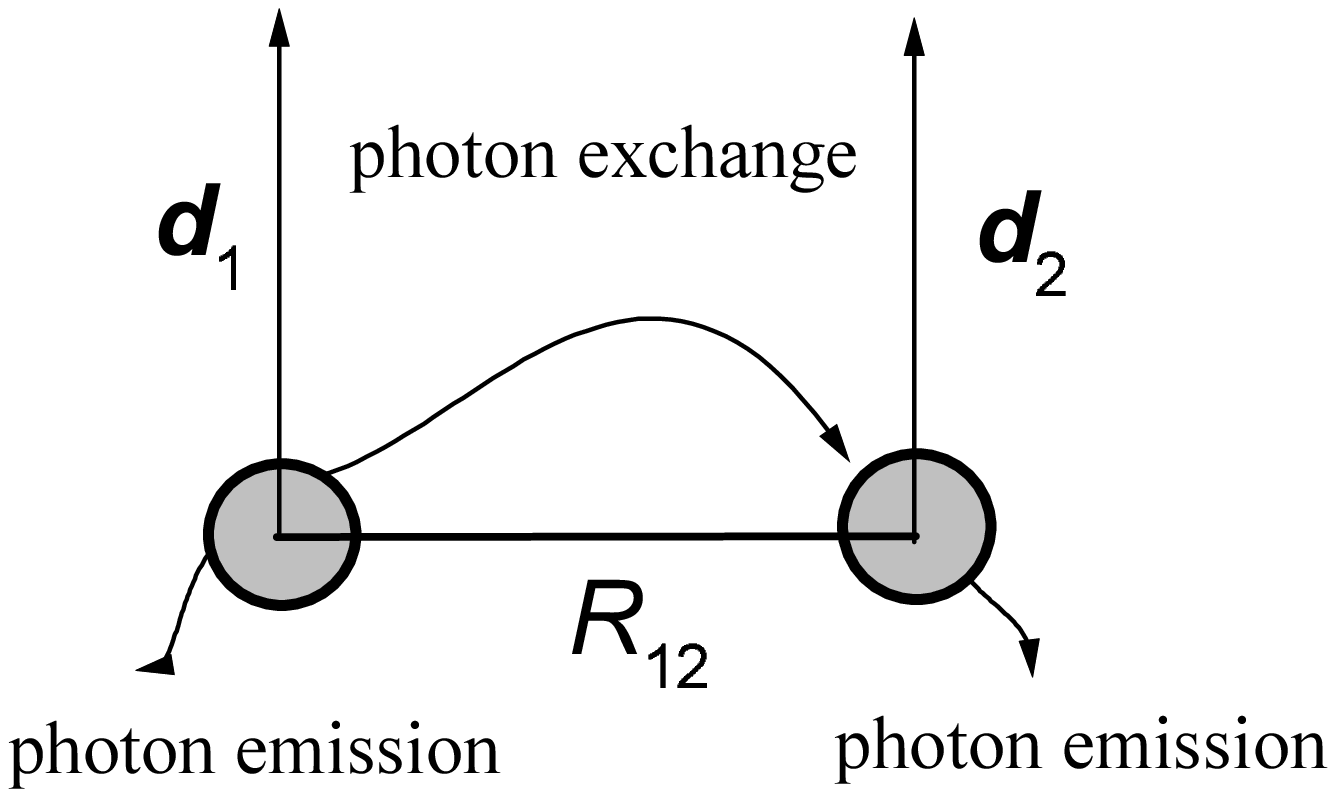}
\end{center}
\caption{Geometry of a TTLA system and relaxation processes of
natural radiative decay.}
\label{system}
\end{figure}

\section{Radiative decay in a system of two closely spaced atoms}
\label{rdecay:}

The system under consideration (Fig. 1) consists of two TLAs
undergoing spontaneous radiative decay of excited states through
the interaction with an electro-magnetic vacuum, which is described
in a standard manner as a thermal reservoir. The relaxation
superoperator ${\cal L}_{\rm r}$ of radiative decay can be
calculated for a system of closely spaced atoms in the same way as
in the case of a single atom [12], i.e., with a standard formula of
second-order perturbation theory in the Hamiltonian of the
interaction of an atom with positive- and negative- frequency
components $\hat\xi^\pm_{k\tau}=-{\bf d}^k_{12}\hat{\bf
E}^\pm_0({\bf R}_k, \tau)$ of (${\bf R}_k$ are the coordinates of
atoms) produced by emission and absorption of photons of the vacuum
electromagnetic field $\hat{\bf E}_0$. Representing superoperators
with the use of a substitution symbol $\odot$ of an operator being
transformed, we arrive at the following expression for the
superoperator ${\cal L}_{\rm r}$ for several atoms in the
Heisenberg representation (by analogy with the case of a single
atom [13])\footnote{In the Schr\"odinger representation with a
properly introduced Hilbert space of atomic operators, the
corresponding operator is described by the superoperator ${\cal
L^+}$ Hermitian-conjugate of ${\cal L}$.}
\begin{eqnarray}
\label{Lr}
{\cal L}_{\rm r}&=&-\lim\limits_{\Delta\to0}
\frac{1}{\hbar^2\Delta}\int\limits_0^\Delta
\int\limits_0^{\tau_2} d\tau_1d\tau_2
\avr{\left[\sum\limits_k \left(\hat\sigma^+_k\hat\xi^-_{k\tau_1}
+\hat\sigma^-_k \hat\xi^+_{k\tau_1}\right),\left[\sum\limits_m
\left(\hat\sigma^+_m\hat\xi^-_{m\tau_2}+\hat\sigma^-_m
\hat\xi^+_{m\tau_2}\right), \odot\right]\right]}_\xi \\ \nonumber
&=&\sum {\cal L}_{{\rm r}k} + \sum_{k\ne m}{\cal L}_{{\rm r}km}.
\end{eqnarray}

\noindent
Here, ${\cal L}_{{\rm r}k}$ is the relaxation eigensuperoperator
for an isolated $k$th atom due to spontaneous decay and ${\cal
L}_{{\rm r}km}$ describe relaxation photon exchange between the
$k$th and $m$th atoms,
\begin{equation}\label{relax}
{\cal L}_{{\rm r}km}=-\frac{\gamma_{km}}{2}\left(\hat\sigma^-_k
\hat\sigma^+_m\odot+\odot\hat\sigma^-_k\hat\sigma^+_m-\hat
\sigma^-_k\odot\hat\sigma^+_m-\hat\sigma^-_m\odot
\hat\sigma^+_k\right).
\end{equation}

\noindent
Here $\hat\sigma^+_k=\hat P^k_{12},$ $\hat\sigma^-_k=\hat
P^k_{21}$, and $\gamma_{kl}$ is the relevant transition rate,
\begin{equation}
\gamma_{kl}=\frac{\omega_{\rm a}^3}{2\pi\hbar c^3}\int
{\bf d}_\bot^k\cdot{\bf d}_\bot^m \exp\left(-i\omega_{\rm a}
{\bf n}{\bf R}_{km}/c\right)d^2{\bf n},
\end{equation}

\noindent where $\bf n$ is the vector along the direction of
the emitted photon, ${\bf d}_\bot^k$ is the transverse
component of the dipole moment, $\omega_{\rm a}$ is the
frequency of atomic transitions, and ${\bf R}_{km}$ are the
vectors of interatomic distances. In the case of two TLAs,
along with the natural rate of radiative decay $\gamma$ for
each atom, for the angle $\theta$ between
parallel dipole moments ${\bf d}^{1,2}_{12}$ and ${\bf
R}_{12}$ equal to $\pi/2$, we obtain a new relaxation
constant of photon exchange $\gamma_{12}=\gamma_{\rm ex}$:
\begin{equation}
\gamma_{\rm ex}=g\gamma, \quad g=\frac{3}{2}\frac{\varphi\cos
\varphi-\sin\varphi+\varphi^2\sin\varphi}{\varphi^3},
\end{equation}

\noindent where the modulus of the parameter $g$ is less than
unity, $\varphi= \omega_{\rm a} R_{12}/c$ is the phase delay of
a signal between the atoms. For $\theta=\pi/2$ and $R_{12}\to0$, we find that
$g\approx1-\varphi^2/5\to1$, whereas in the case of
anti-parallel dipole moments ${\bf d}^{1,2}_{12}$, we have
$g\to-1$. Obviously, these two cases are equivalent to each
other from the information point of view.

With the above-specified assumptions, the noise Liouvillian ${\cal
L}_{\rm r}$ is invariant with respect to the permutation of atoms
and, correspondingly, has no matrix elements that would couple
symmetric and antisymmetric operators of an atomic system.
Therefore, the representation basis is chosen in such a manner that
$\hat e_k=\left(\hat s^1_k\otimes\hat s^2_k\right)/\sqrt{2},
\;k=1,...,4;\quad \hat e_{k+4}=\left(\hat s^1_{i(k)}\otimes\hat
s^2_{j(k)}+\hat s^1_{i(k)}\otimes\hat s^2_{j(k)}\right)/\sqrt{2},\;
k=1,...,6$, and $\hat e_{k+10}=\left(\hat s^1_{i(k)}\otimes\hat
s^2_{j(k)}-\hat s^1_{i(k)}\otimes\hat s^2_{j(k)}\right)/
\sqrt{2},\; k=1,...,6,$ where a pair of indices $i(k)$ and
$j(k)$ describes all possible combinations of pairs of
single-atom basis vectors $\hat s^1_i$ and $\hat s^2_j$:
$\{i(k)j(k)\}=(12,13,14,23,24,34)$. The vectors $\hat s^1_i$
and $\hat s^2_j$ are chosen as a basis of single-atom operators
$\hat I/\sqrt{2}$, $(\hat P_{22}-\hat P_{11})/\sqrt2$, $(\hat
P_{12}+\hat P_{21})/\sqrt2$, and $i(\hat P_{12}-\hat P_{21})/
\sqrt2$, orthonormalized with respect to a scalar product
$(\hat A, \hat B)={\rm Tr}\, \hat A^+\hat B$. Thus, the first
ten basis elements describe symmetric atomic variables, whereas
the last six basis elements describe antisymmetric atomic
variables. Since, in the considered geometry, a laser field
generates only symmetric excitations, the calculation of laser
excitation is reduced to a ten-dimensional
problem.\footnote{For $g=1$, the dimensionality of the problem
can be reduced to 9, i.e., to the dimensionality of a
three-level quantum system that can be obtained from the
starting four-level system by the exclusion of antisymmetric
states. Generally, excitation with a vacuum leads to a random
violation of the symmetry of wave functions but does not change
this symmetry at the level of averaged fluctuations expressed
in terms of the density matrix.} However, this paper does not
consider a procedure of the preparation of the initial state.
We are interested only in the transformation of the initial
state of the form $\hat\rho_i=
\hat\rho^{(1)}\otimes\hat\rho^{(2)}$ in the process of
relaxation.

Applying superoperator (\ref{Lr}) to the basis elements $\hat e_k$,
we obtain the following matrix representation for the relaxation
operator:
\begin{equation}
\label{lr} L_{\rm r}=\left(\begin{array}{cc} L_{\rm SS}&\hat0\\
\hat0&L_{\rm AA}
\end{array}\right)\,,
\end{equation}

\noindent where $\hat 0$ stands for a matrix with zero elements that
complements the remaining submatrices up to a 16$\times$16
matrix, and
$$
L_{\rm SS}=\left(\begin{array}{cc}L_{\rm SS}^1&
0\\ 0&L_{\rm SS}^2\end{array}\right),$$ $$L_{\rm SS}^1=\gamma
\left(\begin{array}{ccccc} 0&0&0&0&-\sqrt{2}\\ 0&-2&g&g&0\\
0&g&-1&0&-\displaystyle\frac{g}{\sqrt{2}}\\
0&g&0&-1&-\displaystyle\frac{g}{\sqrt{2}}\\
0&-\sqrt{2}&\displaystyle\frac{g}{\sqrt{2}}&\displaystyle
\frac{g}{\sqrt{2}}&-1\end{array}\right),\quad L_{\rm
SS}^2=\gamma\left(\begin{array}{ccccc}
-\displaystyle\frac{1}{2}&0&-1-\displaystyle\frac{g}{2}&0&0\\
0&-\displaystyle\frac{1}{2}&0&-1-\displaystyle\frac{g}{2}&0\\
\displaystyle\frac{g}{2}&0&-\displaystyle\frac{3}{2}-g&0&0\\
0&\displaystyle\frac{g}{2}&0&-\displaystyle\frac{3}{2}-g&0\\
0&0&0&0&-1\end{array}\right)
$$

\noindent and
$$
L_{\rm AA}=\gamma\left(\begin{array}{cccccc}
-1&0&0&0&0&0\\
0&-\displaystyle\frac{1}{2}&0&-1+\displaystyle\frac{g}{2}&0&0\\
0&0&-\displaystyle\frac{1}{2}&0&-1+\displaystyle\frac{g}{2}&
0\\
0&-\displaystyle\frac{g}{2}&0&-\displaystyle\frac{3}{2}+g&0&0\\
0&0&-\displaystyle\frac{g}{2}&0&-\displaystyle\frac{3}{2}+g&0\\
0&0&0&0&0&-1
\end{array}\right)
$$

\noindent describe relaxation matrices for symmetric (S) and
antisymmetric (A) variables. Expression (5) shows that these
variables decay independently of each other, and small
eigenvalues $\lambda_k\ll\gamma$ may exist only for the $L_{\rm
SS}$ matrix.

\section{Determination of the amount of quantum information}
\label{concepts:}

The final state at the moment of time $t$ is described by a density matrix
\begin{equation}
\label{Srho}
\hat\rho_f={\cal S}^+(\hat\rho_i^{(1)}\otimes\hat\rho_i^{(2)})\,,
\end{equation}

\noindent
where $\cal S^+$ is the superoperator Hermitian-conjugate of the
superoperator ${\cal S}=\exp\left({\cal L}_{\rm r}t\right)$, which
describes the transformation of operators of physical variables in
the Heisenberg representation. Here, we take into account two pairs
of groups of dynamic variables---variables $\hat A_1$ and $\hat A_2$
related to atoms in the initial state at the moment of time $t=0$
before the interaction and the same variables at a certain final
moment of time $t=\tau$ after the interaction, which determines
transform $\cal S$. From the viewpoint of data exchange, it is of
special interest to consider correlations between atomic variables
at different moments of time, 0 and $\tau$, rather than
correlations between variables $\hat A_1(\tau)$ and $\hat
A_2(\tau)$ at the same moment of time, which are described by
expression (\ref{Srho}). We can calculate the information both
between atomic TLA variables $\hat A_1(0)$ and $\hat A_2(\tau)$ and
between TLA pairs $\hat A_1(0)\,,\hat A_2(0)$ and $\hat
A_1(\tau)\,,\hat A_2(\tau)$.

To define the corresponding amount of information, we introduce the
most typical class of measurement procedures that require the
calculation of the relevant probability distributions. Consider a
standard procedure of measurements with a coincidence scheme, which
is described by a superoperator of the form ${\cal D}^+=
\hat\pi_n^+ \cdot \cdot\cdot\hat\pi^+_1\odot\hat\pi^-_1\cdot\cdot\cdot
\hat \pi^-_n$, where $\hat\pi^\pm_k$ are some Hermitian-field conjugate pairs of
operators, in particular, operators of creation and annihilation.
Suppose that a set of operators $\hat\pi^\pm_m(k)$, $k=1,\ldots,n$
is defined for each $m$ such that the condition of completeness
$\sum_k\hat\pi^-_m(k)\hat\pi^+_m(k)=\hat I$ is satisfied. In fact,
it is of interest to consider sets of eigenprojectors corresponding
to transitions between $n$ eigenstates $\psi_k,\; \hat
A_m\psi_k=\lambda_k\psi_k$, in accordance with relations
$\hat\pi^+_m(k)=\ket{\psi_{k+1}}\bra{\psi_k}$, where $n+1\to1$ is
taken for the brevity of notation. Here, we choose $\hat A_m=\hat
n_m$ for a single atom, where $\hat n_m$ describes an operator of
the number of an energy state at different moments of time
$\tau_m$. For TLAs, we have $\hat\pi^+_1=\ket{2} \bra{1},\;
\hat\pi^+_2= \ket{1} \bra{2},\; \hat\pi^-_1=\hat\pi^+_2$,
and $\hat\pi^-_2=\hat\pi^+_1$. Now, we can introduce the relevant
joint probability distribution,
\begin{equation}
\label{joint}
P(k_1,...,k_n)=\avr{{\cal D}^+(k_1,...,k_n)\hat\rho},
\end{equation}

\noindent
where the superoperator ${\cal D}^+(k_1,...,k_n)=\hat\pi^+_n(k_n)
\cdot\cdot\cdot\hat\pi^+_1(k_1)\odot\hat\pi^-_1(k_1)\cdot\cdot
\cdot\hat\pi^-_n(k_n)$ introduces the corresponding measurement
procedure, and averaging is performed
over both TTLA and reservoir variables with allowance
for the temporal evolution of the system under study
within time intervals $\tau_{m+1}-\tau_m$ between the moments of
time $\tau_m$ corresponding to the observation of detectable
described by the set $\hat\pi^\pm_m$. A formula
$$\hat{\cal E}(k_1,...,k_n)={\cal D}(k_1,...,k_n)\hat I $$

\noindent
relates the superoperator ${\cal D}^+(k_1,...,k_n)$ to a positive
probability operator measure (POM) \cite{QCM}, which is conventionally
employed in the quantum theory of information, i.e., a
nonorthogonal expansion of unity $\hat{\cal E}(d\lambda),
\; \int\hat{\cal E}(d\lambda)=\hat I$, which was introduced
earlier in the theory of optimal quantum solutions (measurements)
for the description of the procedure of a quantum measurement [14].

Here, we will use a set $\hat\pi^\pm_1$ that corresponds to the
populations of atomic levels at the moment of time $\tau_1=0$ and
an analogous set $\hat\pi^\pm_2$ corresponding to the moment of
time $\tau_2$ such that the inequalities
\begin{equation}
\label{tau}
\gamma\tau_2\gg 1\gg(1-g)\gamma\tau_2\,,
\end{equation}

\noindent
are satisfied. In other words, we assume that the distance between
the atoms is sufficiently small, so that the condition $1-g\ll1$ is
met. Then, if inequalities (\ref{tau}) are satisfied for a
transient superoperator ${\cal S}=
\exp({\cal L}_{\rm r} \tau_2)$ we can employ the following approximation:
\begin{equation}
\label{Sg1}
{\cal S}=\lim\limits_{\tau_2\to\infty}
\exp({\cal L}_{\rm r}\mid_{g=1}\tau_2)\,.$$
\end{equation}

\noindent
Under these conditions, the relaxation process brings a
TTLA system into a stationary state, where the decay
rate is equal to the rate of photon exchange, $g\gamma=\gamma$. In
accordance with (8), this approximation is applicable
on a time scale not greater than
$$\tau=[(1-g)\gamma]^{-1}\approx\frac{4\pi^2}{5}
\left(\frac{R_{12}}{\lambda}\right)^2\gamma^{-1}\,,
$$

\noindent
where $\lambda$ is the radiation wavelength for transitions in
TLAs.

Thus, we derive the following averaging rule for an operator of
relaxation dynamics $\cal S$ with $n=2$ in terms of the Heisenberg
representation in the case when quantum numbers $k_1$ of the first
atom are measured at $t=0$ and analogous numbers $k_2$ of the
second atom are measured at $t=\tau$:
\begin{equation}
\label{joint2}
P_{12}(k_1,k_2)={\rm Tr}\hat\rho\hat\pi^-_1(k_1){\cal S}\left[\hat
\pi^-_2 (k_2)\hat\pi^+_2(k_2)\right]\hat\pi^+_1(k_1)\,.
\end{equation}

\noindent
With $n=4$, when the quantum numbers $k_1$ and $k_2$ of the first
and second atoms are measured at the moment $t=0$ and the numbers
$k_3$ and $k_4$ of the same atoms are measured at $t=\tau$, we have
\begin{equation}
\label{joint4}
P_{(12)(34)}(k_1,k_2,k_3,k_4)={\rm Tr}\hat\rho\hat\pi^-_1(k_1)
\hat\pi^-_2(k_2) {\cal S}\left[\hat\pi^-_1(k_3)\hat\pi^-_2 (k_4)
\hat\pi^+_2(k_4)\hat\pi^+_1(k_3)\right]\hat\pi^+_2(k_2)\hat\pi^+_1
(k_1)\,.
\end{equation}

\noindent
The amounts of information corresponding to these distributions are
written as
\begin{equation}
\label{i12}
I_{12}=H(P_1)+H(P_2)-H(P_{12})
\end{equation}

\noindent
in the case of separate atoms and
\begin{equation}
\label{i1234}
I_{(12)(34)}=H(P_{(12)})+H(P_{(34)})-H(P_{(12)(34)})
\end{equation}

\noindent
for pairs of atoms. Here, we employed a functional of
the entropy $H(w)=-\sum \limits_kw(k)\log_2w(k)$ and
single-moment probability distributions
$$
P_1(k)=\sum\limits_{k_2}P_{12}(k,k_2),\quad
P_2(k)=\sum\limits_{k_1}P_{12}(k_1, k),
$$
$$
P_{(12)}(k_1,k_2)=\sum\limits_{k_3,k_4}P_{(12)(34)}
(k_1,k_2,k_3, k_4),\quad P_{(34)}(k_3,k_4)=
\sum\limits_{k_1,k_2}P_{(12)(34)}(k_1,k_2,k_3,k_4).
$$

\section{Results of calculations}
\label{calc:}

In the absence of relaxation, i.e., with ${\cal S}=1$, the final
state (\ref{Srho}) coincides with the initial state,$\hat\pi^\pm_2=
\hat\pi^\pm_1$, and joint distributions (\ref{joint2}) and (\ref{joint4})
can be represented as products of two independent distributions of
the form $P_1(k_1) P_2(k_2)$ and $P_1(k_1)P_2(k_3)
\delta(k_1-k_2)\delta(k_3-k_4)$, respectively, where $P_l(k)= {\rm
Tr}\,\hat\pi^+_1(k)\hat\rho_i^{(l})\hat\pi^-_1(k)$. Obviously, the
corresponding amounts of information described by (\ref{i12}) and
(\ref{i1234}) are equal to zero in this case, i.e., the
establishment of information occurs only in the process of
relaxation.

Since we consider information in a quasi-stationary state, we have
to calculate the transient superoperator (\ref{Sg1}) neglecting the
decay of a single-quantum pure state $\psi_{\rm a}$, which occurs
with a rate equal to $(1-g)\gamma$. To perform such a calculation,
we use the following representation:
$$
{\cal S}=\lim\limits_{\lambda\to0}\lambda
\left({\cal L}_{\rm r}-\lambda I\right)^{-1},
$$

\noindent
where $I$ describes the identity superoperator represented by an
identity matrix of the corresponding dimensionality, which is equal
to 16 in the case of TTLAs. The resulting matrix representation
${\cal S}$ with $g=1$ is written as
\begin{equation}
\label{S0}S_0 = \left(\begin{array}{cccccc}
1&\displaystyle\frac{1}{2}&-\displaystyle\frac{1}{4}&
-\displaystyle\frac{1}{4}&
-\displaystyle\mathstrut\frac{3}{2\sqrt{2}}&\hat0\\
0&\displaystyle\frac{1}{2}&\displaystyle\frac{1}{4}&\displaystyle
\frac{1}{4}&-\displaystyle\mathstrut\frac{1}{2\sqrt{2}}&\hat0\\
0&\displaystyle\frac{1}{2}&\displaystyle\frac{1}{4}&\displaystyle
\frac{1}{4}&-\displaystyle\mathstrut\frac{1}{2\sqrt{2}}&\hat0\\
0&\displaystyle\frac{1}{2}&\displaystyle\frac{1}{4}&\displaystyle
\frac{1}{4}&-\displaystyle\mathstrut\frac{1}{2\sqrt{2}}&\hat0\\
& & &\hat0& &\hat0\end{array}\right).
\end{equation}

\noindent
For $g<1$, we derive an expression
$$
S=\left(\begin{array}{cccccc}1&1&0&0&-\sqrt{2}& \hat 0\\
&&&\hat 0&&\hat 0\end{array}\right),
$$

\noindent
which takes into account the decay of all the atomic states except
for the ground one. The density matrix of the stationary state in
this case is written as
\begin{equation}
\hat\rho^{(1+2)}(t)={\cal S}^+\hat\rho^{(1+2)}=
\left(\begin{array}{cc}1&\hat0\\ \hat0&\hat0
\end{array}\right).
\end{equation}

\noindent
Such a density matrix corresponds to a quantum-free state
$\psi_0=\ket0\otimes\ket0$, i.e., the ground state of each atom,
and, consequently, contains no information concerning the initial
state of atoms. By contrast, the stationary density matrix
corresponding to expression (14) is written as
\begin{equation}
\hat\rho^{(1+2)}(t)={\cal S}^+\hat\rho^{(1+2)}=
\left(\begin{array}{cccc}1-A&0&0&0\\ 0&A/2&-A/2&0\\ 0&-A/2&A/2&0\\
0&0&0&0\end{array}\right)\,,
\end{equation}

\noindent
where
$$
A=\frac{1}{2}\left(\rho^{(2)}_{22}-\rho^{(2)}_{21}
\rho^{(1)}_{12}-\rho^{(2)}_{12} \rho^{(1)}_{21}+\rho^{(1)}_{22}-
2\rho^{(2)}_{22}\rho^{(1)}_{22}\right).
$$

\noindent
Such a density matrix corresponds to a mixed state of the form
\begin{equation}\hat\rho^{(1+2)}(t)=
(1-A)\psi_0\psi_0^++A\psi_{\rm a}\psi_{\rm a}^+,
\end{equation}

\noindent
In this case, the quantity $A$ represents the probability that a
TTLA system resides in the singlet state $\psi_{\rm a}$ upon the
completion of fast relaxation processes. In fact, off-diagonal
elements of the density matrix with indices 12 and 21 are not
involved in the matrix corresponding to the joint probability
distribution of the quanta of the first and second atoms:
\begin{equation}
\parallel P(k_1,k_2)\parallel =
\left(\begin{array}{cc}\rho^{(2)}_{22}\rho^{(1)}_{22}/4&(4-
\rho^{(2)}_{22})\rho^{(1)}_{22}/4 \\
\rho^{(2)}_{11}\rho^{(1)}_{11}/4&(3+\rho^{(2)}_{22})
\rho^{(1)}_{11}/4 \end{array}\right).
\end{equation}

\begin{figure}
\begin{center}
\epsfxsize=5.cm\epsfclipon\leavevmode\epsffile{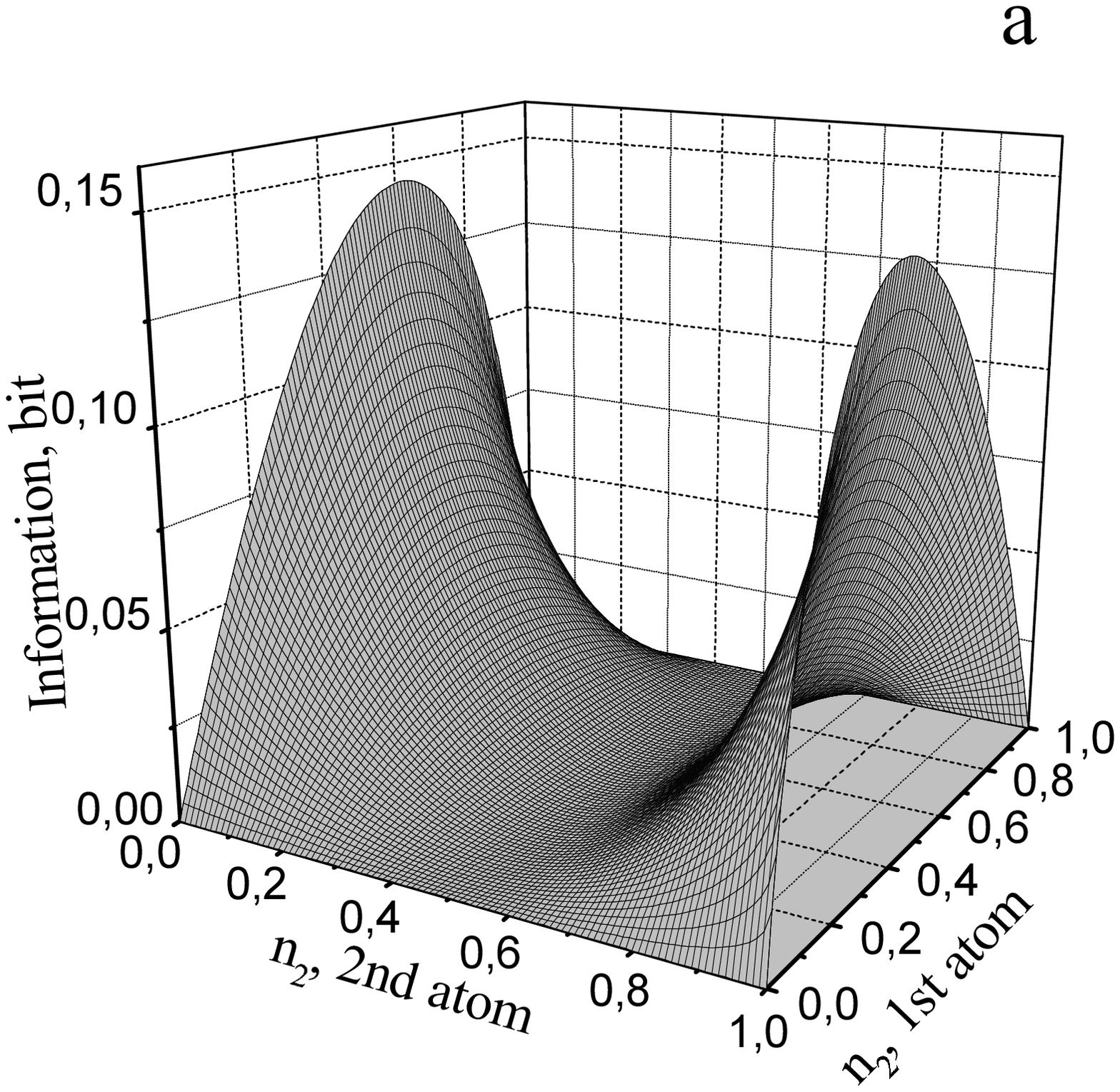}
\epsfxsize=5.5cm\epsfclipon\leavevmode\epsffile{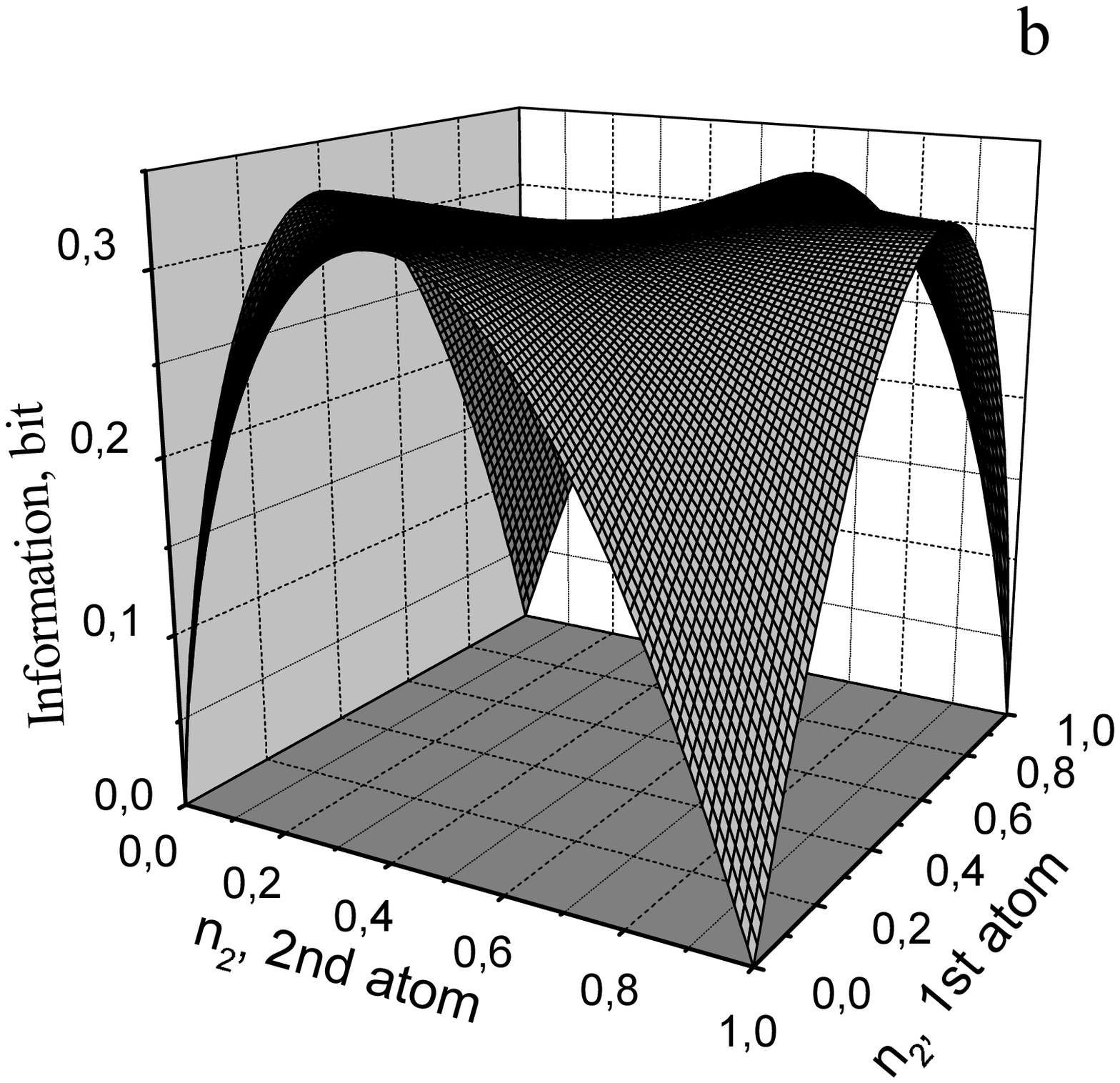}
\end{center}
\caption{The amounts of information (a) between the initial state
of the first atom and the final state of the second atom and
(b) between the initial and final states of both atoms as functions
of the mean populations of excited levels of the initial atomic states.}
\label{info}
\end{figure}

\noindent
For $g=-1$, we obtain exactly the same distribution. The
corresponding amount of information (\ref{i12}) is presented in
Fig. 2a. The maximum of the information obtained between the final
state of the second atom and the initial state of the first atom is
achieved with a population $n_2=0$ or $n_2=1$ for the second atom
and $n_2=-1+\sqrt{37}/4=0.521$ or, respectively, $n_2=2-\sqrt{37}/4
=0.479$ for the first atom. The maximum amount of information is
given by
$$
\begin{array}{c}
I_{12}^{\rm max}=\left(\displaystyle\frac{3}{2}-\displaystyle\frac{3\sqrt{37}}{16}\right)
\log_2\left(\displaystyle\frac{3}{2}-\displaystyle\frac{3\sqrt{37}}{16}\right)-
\left(2-\displaystyle\frac{\sqrt{37}}{4}\right)\log_2\left(2-\displaystyle\frac{\sqrt{37}}
{4}\right)-\left(\displaystyle\frac{1}{2}+\frac{\sqrt{37}}{16}\right)\log_2
\left(\displaystyle\frac{1}{2}+\frac{\sqrt{37}}{16}\right),
\end{array}
$$

\noindent
and is approximately equal to 0.14 bit.

Joint two-atom two-moment distribution (\ref{joint4}) is
represented by a matrix
\begin{equation}
\parallel P(\{k_1,k_2\},\{k_3,k_4\})\parallel =
\left(
\begin{array}{cccc}
0&0&0&\rho^{(2)}_{11}\rho^{(1)}_{22}/4\\ 0&\rho^{(2)}_{22}
\rho^{(1)}_{22}&\rho^{(2)}_{11} \rho^{(1)}_{22}/4&\rho^{(2)}_{11}
\rho^{(1)}_{22}/2\\ 0&\rho^{(2)}_{22}\rho^{(1)}_{11}/4&0&0\\
\rho^{(2)}_{22} \rho^{(1)}_{11}/4&\rho^{(2)}_{22}
\rho^{(1)}_{11}/2&0&\rho^{(2)}_{11}\rho^{(1)}_{11}
\end{array}\right).
\end{equation}

\noindent
The corresponding amount of information (\ref{i1234}) in both atoms
at the moment $t=\tau$ as compared with the initial state is
presented in Fig.\ \ref{info}b. The maximum of this quantity is
$I_{(12)(34)}^{\rm max}=0.322$ bit.

\section{Conclusions}

The results of calculations presented above demonstrate that there
exists a natural mechanism of information exchange between two
closely spaced two-level atoms. If these atoms are initially
prepared in statistically independent states, then, after a lapse
of time $\tau$ such that $(1-g)\gamma\tau \ll1\ll\gamma\tau$, where
$\gamma$ is the decay rate of single-atom excitations and
$g\approx1$ is the efficiency of interatomic photon exchange with
respect to single-atom decay, the procedure of counting the quanta
of atomic excitations reveals the presence of nonzero information
between the population of the second atom at the moment of time
$\tau$ and the initial population of the first atom. The maximum
amount of information $I_{\rm max}=0.14$ bit corresponds to the
upper or lower initial state of the second atom and the mean
population of the first atom close to 0.5. An analogous amount of
information in both atoms is equal to $I_{\rm max}=0.322$ bit.
Thus, a certain amount of information is produced without special
efforts due to the specific features of spontaneous radiative decay
in a system of closely spaced atoms.

Taking into account not only the relaxation but also resonant
electrostatic dipole-dipole interaction, which is described by a
Hamiltonian of the form $\hat{\cal H}_C=g_C(\hat\sigma_1^+\hat
\sigma_2^-+\hat\sigma_1^-\hat\sigma_2^+)$, one can easily verify
by direct calculations that the superoperator of the transient
distribution (\ref{S0}) remains unchanged. The same conclusion
follows from the consideration of the symmetry of the system with
allowance for the invariance of $\hat{\cal H}_C$ with respect to
the permutation of atoms. Thus, electrostatic dipole-dipole
interaction has no influence on relaxation data exchange as long as
this interaction remains too weak to change the relaxation
operator.

The results obtained above cannot be directly applied to the
analysis of atoms on a surface because of large nonradiative
dephasing rates, which gives rise to the main difficulty
encountered when attempts are made to employ adsorbed atoms for the
implementation of quantum computations [3]. However, we should note
that, for closely spaced atoms, analysis of dephasing effects
should take into account the specific features of relaxation
processes associated with the interaction of atoms through
dephasing excitations. The mechanism behind such an interaction
should be, in a certain respect, similar to the mechanism
considered in this paper.

\acknowledgments

We are grateful to V.I. Panov for stimulating discussions. This
study was supported in part by the Russian Foundation for Basic
Research (project No. 96-03-32867) and Volkswagen Stiftung (grant
No. 1/72944). V. N. Z. also acknowledges the support of the
Alexander von Humboldt Foundation, Germany.

\end{document}